\documentclass[a4paper,twocolumn,english,aps]{revtex4-1}

\usepackage[T1]{fontenc}
\usepackage[utf8]{inputenc}
\usepackage[english]{babel}
\usepackage{amsmath,amsfonts,amsthm,amssymb}
\usepackage{mathtools}
\usepackage{graphicx}
\usepackage{color}
\usepackage[pdftex]{hyperref}
\usepackage{bm}
\usepackage{natbib}
\usepackage{url}
\usepackage{textcase}
\usepackage{soul}

\newcommand{\ep}{\epsilon}

\begin{document}

\title{Plasmon-exciton polaritons in 2D semiconductor/metal interfaces}

\author{P.~A.~D.~Gon\c{c}alves$^{1,2,3}$, L.~P.~Bertelsen$^{2}$,  Sanshui~Xiao$^{2,3}$, and N.~Asger~Mortensen$^{1,3,4}$}
\email{asger@mailaps.org}

\affiliation{$^{1}$Center for Nano Optics, University of Southern Denmark, Campusvej 55, DK-5230~Odense~M, Denmark}
\affiliation{$^{2}$ Department of Photonics Engineering, Technical University of Denmark, DK-2800 Kgs.~Lyngby, Denmark}
\affiliation{$^{3}$Center for Nanostructured Graphene, Technical University of Denmark, DK-2800 Kgs.~Lyngby, Denmark}
\affiliation{$^{4}$Danish Institute for Advanced Study, University of Southern Denmark, Campusvej 55, DK-5230~Odense~M, Denmark}
\date{\today}

\begin{abstract}
The realization and control of polaritons is of paramount importance in the prospect of novel photonic devices. 
Here, we investigate the emergence of 
plasmon-exciton polaritons in hybrid structures consisting of a two-dimensional (2D) transition metal dichalcogenide (TMDC) 
deposited onto a metal substrate or coating a metallic thin-film. 
We determine the polaritonic spectrum and show that, in the former case, the addition of a top dielectric layer, and, in the latter, the thickness of the metal film,
can be used to tune and promote plasmon-exciton interactions well within the strong coupling regime. 
Our results demonstrate that Rabi splittings exceeding 100\,meV can be 
readily achieved in planar dielectric/TMDC/metal structures under ambient conditions. 
We thus believe that this work provides a simple and intuitive picture to tailor strong coupling in plexcitonics, with potential applications for engineering compact photonic devices with tunable optical properties.
\end{abstract}

\maketitle
% ==============================================================================================
% ::                                    INTRODUCTION                                          ::
% ==============================================================================================
%\section{Introduction}

\textit{Introduction.} Two-dimensional (2D) materials~\cite{novoselov2005two} have been subject of increasing interest in nanophotonics due to their ability to host a 
large variety of polaritons~\cite{Low:2017,Basov:2016}, including gate-tunable plasmon polaritons in graphene~\cite{Goncalves:2016,Fei:2012,Chen:2012,Xiao:2016} 
or phonon polaritons in hexagonal boron nitride (hBN)~\cite{Dai:2014,Caldwell:2014,Caldwell:2015}. More recently, exciton polaritons 
in atomically thin semiconductors such as transition metal dichalcogenides (TMDCs) have also been observed experimentally~\cite{Hill:2015,flatten:2016,hu2017imaging}, thus further expanding the current 2D "polaritonic library". 
%The chemical formula of TMDCs can be written as ${\rm MX}_2$, where ${\rm M}$ labels a 
%transition metal (like Mo or W) and ${\rm X}$ a chalcogen (for instance, S, Se, or Te).

Excitons---electron-hole pairs bounded through their mutual Coulomb interaction---in 2D materials exhibit large binding energies when compared with excitons in conventional 3D semiconductors. This fact is direct consequence of the modification of the Coulomb potential in two dimensions, which results in less screening and thereby higher excitonic binding energies~\cite{wang2017excitons}. Therefore, excitons in 2D TMDCs strongly affect their electromagnetic response and light-matter interactions even at room temperature. For this reason, atomically thin TMDCs are currently emerging as viable platforms for studying strong coupling under ambient conditions. Indeed, strong coupling and mode hybridization of excitons in 2D materials with photonic cavity modes~\cite{liu:2015,flatten:2016} and with plasmonic resonances~\cite{Liu:2016,Wen:2017,Abid:2017} have been recently reported. 
These advances have thus unambiguously placed TMDCs in the map as suitable candidates for emerging quantum technologies~\cite{AsgerQuantum} relying on strong coupling and coherent quantum control, and can be viewed as potential alternatives to (or used in conjunction with) typical systems such as J-aggregates of organic dye molecules or quantum dots. 
In this context, while promoting exciton-photon interactions in photonic cavities may have the advantage of rendering lower losses, this comes at the expense of larger modal volumes when compared with plasmonic systems. 
On the other hand, the latter, despite exhibiting more losses, can deliver strong optical confinement and enhance light-matter interactions at the nanometer scale. This feature is pivotal, for instance, for achieving strong coupling regimes at the single molecule level~\cite{Baumberg:2016}.

\begin{figure}[b]
\centering
 \includegraphics[width=1.0\columnwidth]{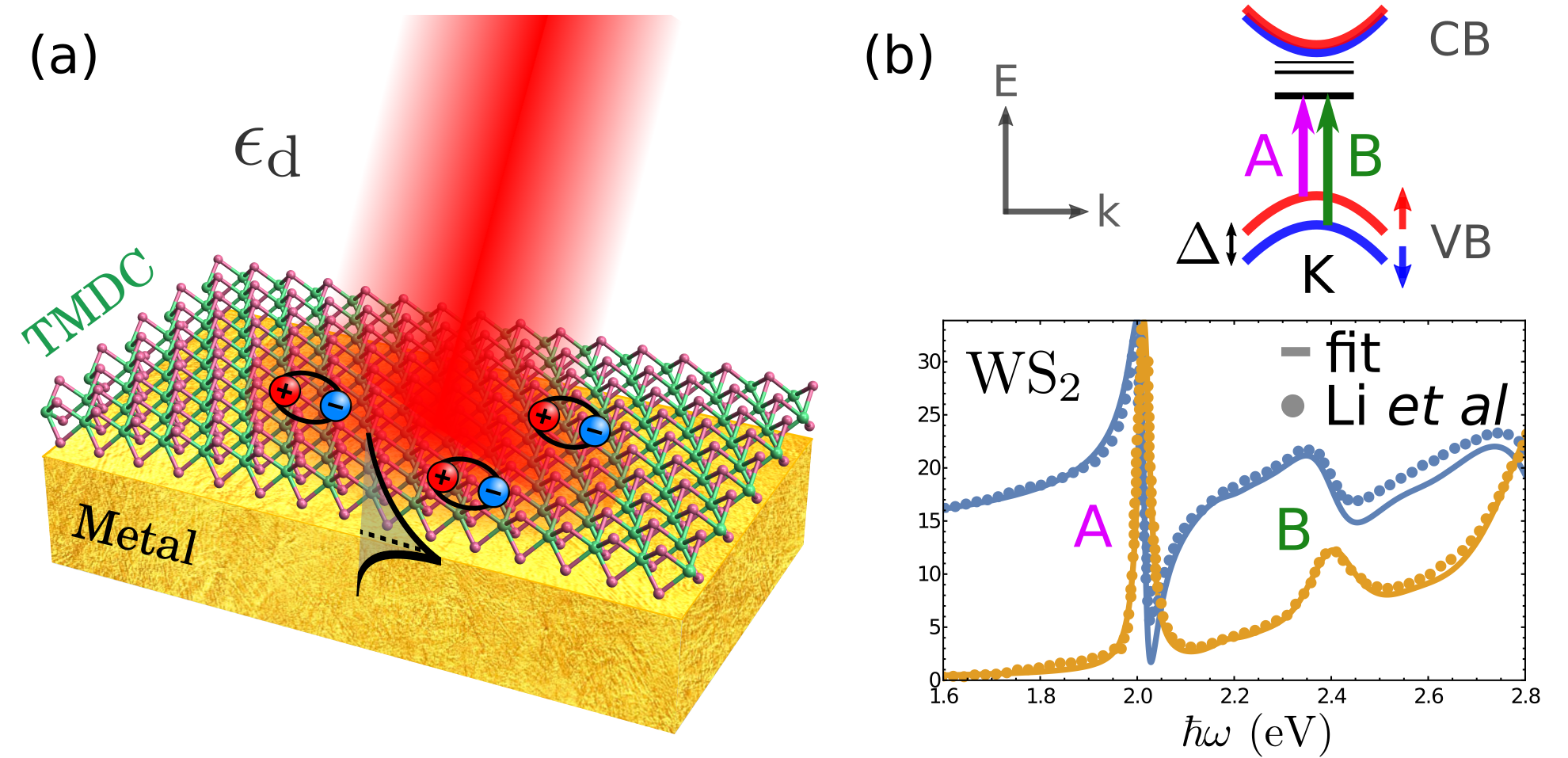}
  \caption{(a) Illustration of the dielectric/TMDC/metal system under consideration in this work. 
  (b) Scheme of a TMDC electronic structure around the $K$ point, showing optical transitions from the 
  spin-orbit split valence band to the exciton ground state (dubbed A and B excitons). 
  Also plotted is the dielectric function data of WS$_2$ measured by Li \textit{et al.}~\cite{Li:2014}, 
  together with the corresponding Lorentz oscillator fit, 
  $\ep_{{\rm WS}_2}(\omega) = \ep_\infty + \sum_j f_j/(\omega^2_j-\omega^2 -i\omega\gamma_j)$ 
  (see endnote \cite{noteEps} for details on the parameters).
  }\label{fig:1}
\end{figure}

In this Rapid Communication, we theoretically investigate plasmon-exciton coupling in planar structures consisting of an atomically thin TMDC crystal deposited onto a metal substrate, 
as depicted in Fig.~\ref{fig:1}(a). Upon resonant excitation, excitons can be created in the TMDC, while the metal can sustain surface plasmon 
polaritons (SPPs). Hence, in a system comprised of two such materials, 
\begin{figure*}[t]
 \centering
  \includegraphics[width=1.0\textwidth]{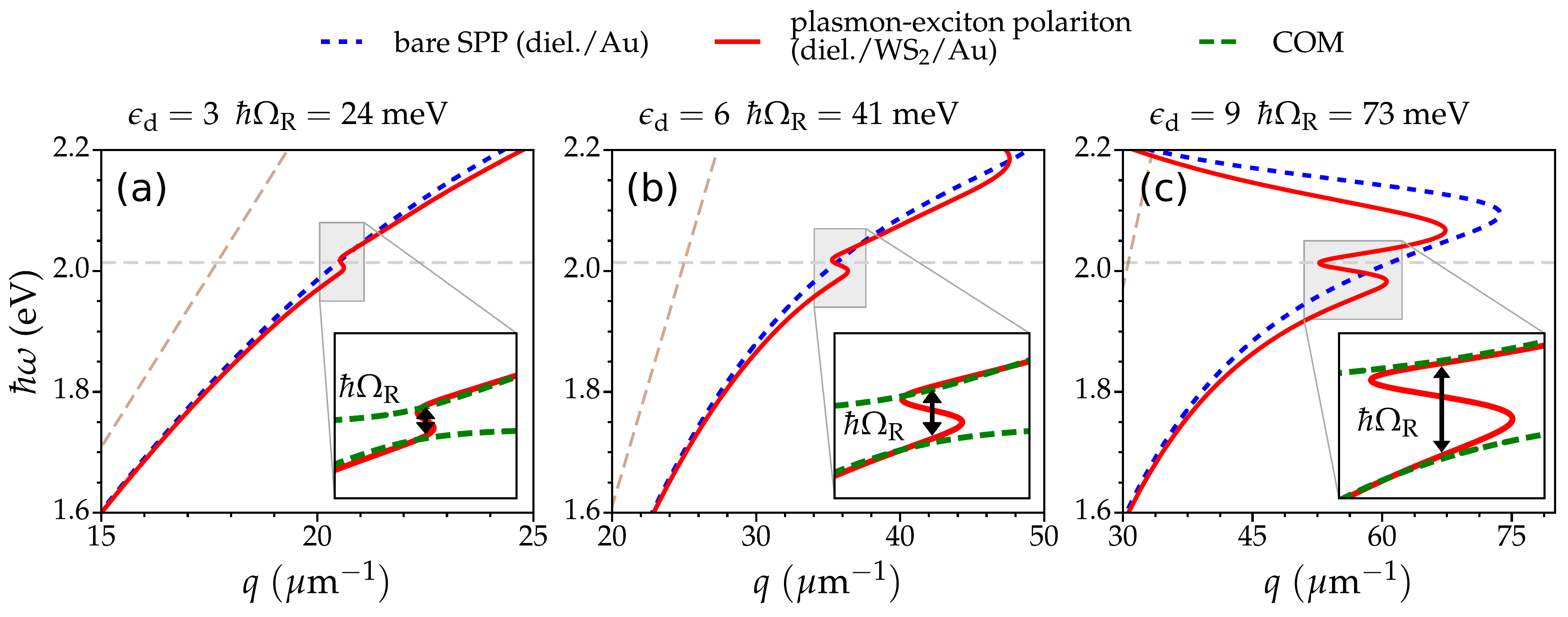}
   \caption{Dispersion relation, $\hbar\omega$ versus ${\rm Re}\{q\}$, of plasmon-exciton polaritons in dielectric/WS$_2$/Au structures (red solid lines) with different 
   top dielectrics, as indicated at the top of each panel. The dispersion of SPP in the corresponding dielectric/Au interfaces 
   is also plotted (blue dashed lines), along with energy of the WS$_2$ exciton A (horizontal dashed line), and the light lines 
   (dashed lines in light brown). The insets in each panel show a zoom of the area around the mode back-bending/anticrossing, together 
   with the Rabi splitting (marked by the arrows). The retrieved Rabi splitting energies are shown at the top of each panel, 
   and were used as a parameter in the analytical coupled oscillator model (COM), whose result (up to minute absolute vertical shift) is depicted by the green dashed 
   lines in the insets.}
   \label{fig:2}
\end{figure*}
the interaction between the individual excitations gives rise to hybrid plasmon-exciton polaritons. When the system is driven into the strong coupling regime, energy can be coherently exchanged between the two eigenstates of the compound system. The signature of this phenomena typically manifests itself in the optical spectra in the form of a Rabi splitting, together with the emergence of anticrossing behavior in the polaritonic dispersion diagram. 
Here, we show that strong coupling between excitons and plasmons can be readily achieved in elementary dielectric/TMDC/metal interfaces by choosing appropriately the dielectric constant of the superstrate, or alternatively, the thickness of a TMDC-coated metal thin-film. 
We compute the full plasmon-exciton polariton spectrum and retrieve the ensuing Rabi splitting, which we then feed into a coupled oscillator model (COM) for comparison. Our results demonstrate that Rabi splittings in excess of 100\,meV can be achieved in such systems at room temperature. 
Finally, as an application of our theoretical results, we consider an experimental setting in which plasmon-exciton polaritons are excited by exploiting the attenuated total reflection (ATR) technique in the so-called Otto configuration~\cite{Goncalves:2016}. The corresponding results plainly show the manifestation of strong coupling in which the two hybrid modes are well separated according to the standard Rabi splitting to linewidth criterion~\cite{torma2014strong}. We believe our work can be used as a springboard to tailor plasmon-exciton interactions in 2D materials interfacing 3D metallic structures, potentially opening new avenues to engineer the interplay between light and matter at the nanoscale.

% ==============================================================================================
% ::                                      MAIN BODY                                           ::
% ==============================================================================================

\textit{Theory and results.} The dispersion relation for polaritons sustained at a dielectric/TMDC/metal interface, like the one portrayed in Fig.~\ref{fig:1}(a), directly follows from the implicit condition~\cite{Goncalves:2016,Xiao:2016}
\begin{equation}
 \frac{\epsilon_{\rm d}}{\kappa_{\rm d}} + 
 \frac{\epsilon_{\rm m}}{\kappa_{\rm m}} 
 = \frac{\sigma_{\mathrm{2D}}}{i \omega \epsilon_0}, 
 \label{eq:dispersion}
\end{equation}
where $\kappa_j^2 = q^2 - \ep_j \omega^2/c^2$ with $j=\{\rm{d}, \rm{m} \}$. Here, $\epsilon_{\rm d}$ stands for the relative permittivity of the upper dielectric, and $\epsilon_{\rm m}(\omega)$ represents the metal's complex dielectric function. The electromagnetic properties of the 2D TMDC are accounted for via the material's (surface) conductivity, $\sigma_{\mathrm{2D}}(\omega)$. 
In what follows, and for the remaining of the Letter, we consider the metal substrate to be made of gold, and the 2D TMDC to be a monolayer of tungsten disulphide (WS$_2$). The reason is that the latter exhibits excitonic resonances with large oscillator strengths~\cite{Li:2014}, while the former is a widely used plasmonic medium. Furthermore, in order to obtain results that are faithful in the most possible way to realistic experimental conditions, we model gold via its experimentally obtained dielectric function~\cite{ExpAu}. 
Similarly, we construct the 2D conductivity of the WS$_2$ monolayer out of the dielectric function data measured by Li \textit{et al.}, to which a Lorentz oscillator has been fitted~\cite{Li:2014}---see also Fig.~\ref{fig:1}(b) for details. In particular, we take 
$\sigma_{\mathrm{2D}}(\omega) = - i \omega \ep_0 d (\epsilon_{\rm TMDC}(\omega)  - 1)$ where $d$ is an effective thickness, typically the material's bulk interlayer spacing ($d_{{\rm WS}_2} = 6.18$\AA{} for WS$_2$)~\cite{Li:2014}. As Fig.~\ref{fig:1}(b) portrays, the WS$_2$ dielectric function is dominated by two resonances attributed to its A and B excitons, corresponding to optical transitions between the spin-orbit split valence band and the $n=1$ exciton state. For a detailed account of the exciton properties in 2D TMDCs, see Ref.~\onlinecite{Thygesen:2017}.

In possession of the materials' response functions, the polaritonic spectrum stems from the numerical solution of Eq.~(\ref{eq:dispersion}) in the $(q,\hbar\omega)$ phase space.
Figure~\ref{fig:2} shows the dispersion diagram of plasmon-exciton polaritons in three dielectric/WS$_2$/Au structures with different capping dielectrics. The bare SPP dispersion in a dielectric/Au interface is also plotted and serves as an eye-guide. 
It is apparent from the figure that plasmon-exciton hybridization arises in the neighborhood where the uncoupled SPP crosses the energy of the A exciton. Since we solve Eq.~(\ref{eq:dispersion}) by feeding in a real frequency, the corresponding numerical solution yields a complex-valued $q$ (whose real part is plotted in Fig.~\ref{fig:2}), 
and thus a back-bending or anomalous dispersion behavior emerges rather then the usual anticrossing~\cite{Goncalves:2016,hu2017imaging} 
%and thus the usual anticrossing emerges instead as a characteristic back-bending or anomalous dispersion behavior
(the more familiar anticrossing presents itself in experiments at fixed angles and varying excitation energies; we shall consider this case below). We note that such a back-bending is an unambiguous signature of strong coupling~\cite{Wolff:2017}. 
Additionally, the inset in each panel in Fig.~\ref{fig:2} displays zoomed versions of the area where such mode hybridization occurs. The span of the insets' vertical axes is the same to facilitate the comparison of the Rabi splitting between panels. 
Figure~\ref{fig:2} clearly shows that the Rabi splitting energy (indicated by the arrows) significantly increases as the dielectric constant of the cladding dielectric gets larger. Incidentally, our calculations have shown 
that for an air/WS$_2$/Au planar structure the Rabi splitting is almost absent. However, Fig.~\ref{fig:2} demonstrates that the addition of a dielectric top layer can be used to substantially enlarge its value. In fact, the Rabi splitting energy increases from a barely tangible $\hbar\Omega_{\rm R} = 24$\,meV, for $\ep_{\rm d}=3$ in Fig.~\ref{fig:2}(a), to a striking value of $\hbar\Omega_{\rm R} = 73$\,meV, for $\ep_{\rm d}=9$ in Fig.~\ref{fig:2}(c). 
With the purpose of both validating the numerically retrieved Rabi splittings, and in the interest of providing physical insight, we have introduced these as parameters on the coupled oscillator model (COM), which is widely adopted in the literature to describe strong coupling~\cite{torma2014strong,Liu:2016,Wen:2017,flatten:2016}. Within the COM framework, the frequencies of the two hybrid modes read (neglecting damping to simplify the analysis)
\begin{equation}
\omega_\pm (q) = \frac{\omega_{\rm pl}(q) + \Omega_{\rm ex}}{2} \pm \frac{1}{2} \sqrt{ [\omega_{\rm pl}(q) - \Omega_{\rm ex}]^2 + \Omega_{\rm R} ^2 },
 \label{eq:COM}
\end{equation}
where $\omega_{\rm pl}(q)$ 
refers to the bare SPP dispersion [obtained from Eq.~(\ref{eq:dispersion}) with $\sigma_{\mathrm{2D}}\rightarrow 0$], and $\hbar\Omega_{\rm ex}$ is the A exciton energy. The comparison of the plasmon-exciton dispersion and corresponding $\hbar\Omega_{\rm R}$ determined numerically with the analytical results from the COM reveals an outstanding agreement, 
as can be seen from the insets in Fig.~\ref{fig:2}.
\begin{figure}[t]
\centering
 \includegraphics[width=0.7\columnwidth]{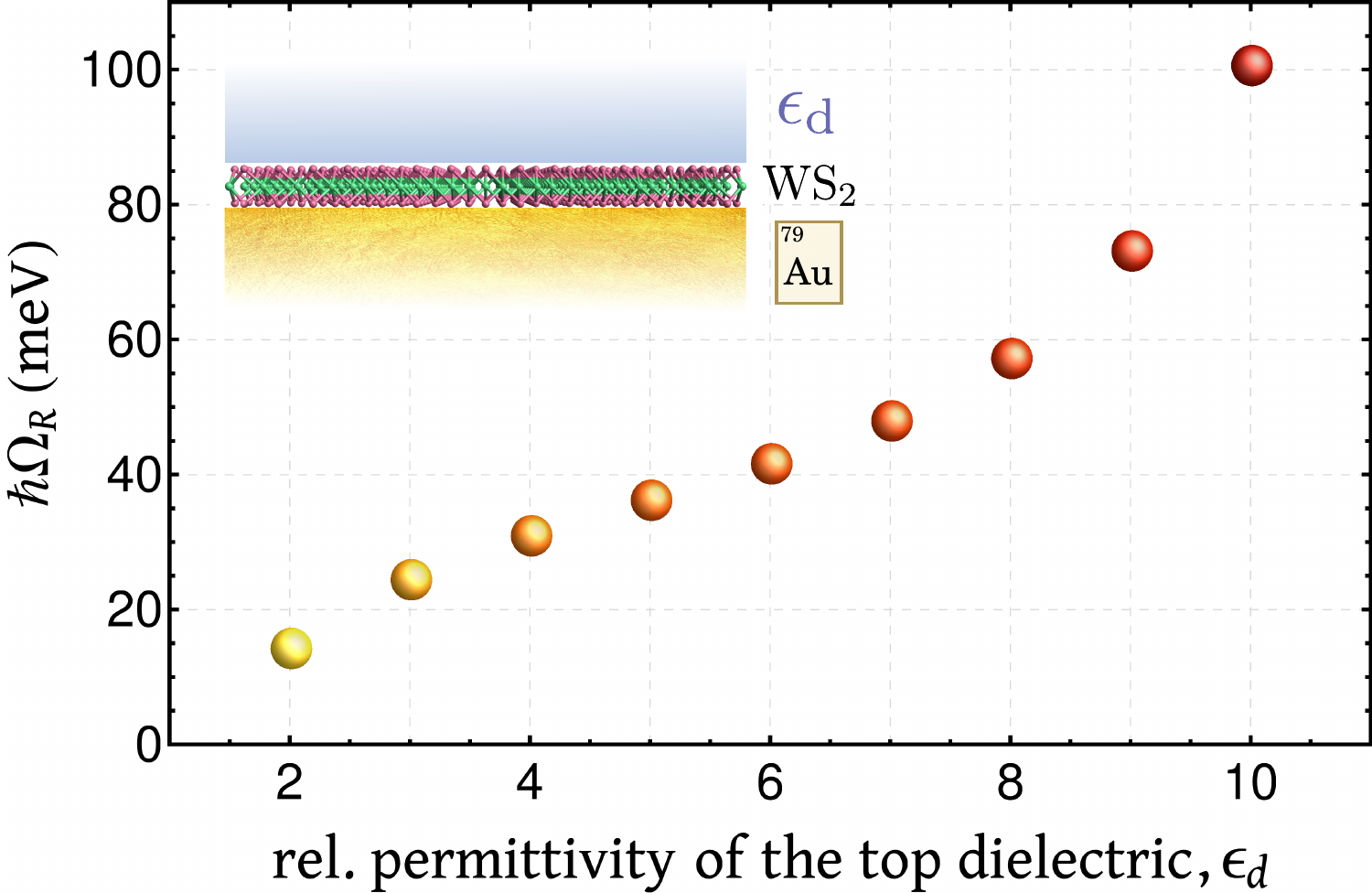}
  \caption{Rabi splitting energy (in meV) as a function of the dielectric constant of the top insulator in dielectric/WS$_2$/Au structures.}
  \label{fig:3}
\end{figure}
Augmenting the previous results to a larger set of dielectric claddings, one can construct a map of the Rabi splitting as a function of $\ep_{\rm d}$. The outcome of such a procedure is shown in Fig.~\ref{fig:3}. As discussed above, the Rabi splitting energy increases monotonically upon increasing the relative permittivity of the top dielectric medium (see Supplemental Material~\cite{SuppMat} for a detailed analysis).
These results can be understood by noting that, for a Drude metal (for the sake of clarity), the nonretarded surface plasmon frequency is given by $\omega_{\rm sp}=\omega_{\rm p}/\sqrt{1+\ep_{\rm d}}$. In typical metals, such resonance lies well above the energy of excitons in TMDCs, and consequently the corresponding plasmon-exciton coupling is weak. Yet, as one increases the permittivity of the dielectric, $\omega_{\rm sp}$ undergoes a redshift and approaches the excitonic resonance, thereby promoting successively stronger plasmon-exciton interactions.
Accordingly, the corresponding mode hybridization also occurs at larger $q$'s, so that a 
higher proportion of the field distribution lies in the vicinity of the 2D excitonic medium (i.e., the spatial overlap between the field of the SPP and the exciton increases).
In previous experiments, a similar tuning is carried out by depositing metallic nanoparticles (NPs) on the 2D material, so that the plasmon resonance is alternatively controlled by changing the NPs' size~\cite{Liu:2016,Wen:2017,Abid:2017}. 
The advantage of our approach is that it eliminates the necessity of nanostructuring. 
Our results therefore demonstrate the ability to achieve plasmon-exciton strong coupling, at room temperature, in simple and highly scalable dielectric/TMDC/metal configurations.
We now consider an application of our results, namely the excitation of plasmon-exciton polaritons via prism coupling using the ATR technique. In this scenario, the system's polaritons can be excited by the evanescent waves coming out of a prism placed on top of the dielectric/WS$_2$/Au structure (i.e., photons can ``tunnel'' from the prism to the WS$_2$/Au interface through the dielectric layer). 
All our calculations relative to the ATR configuration were performed using the transfer-matrix method (see Ref.~\cite{Goncalves:2016} for more information on this technique). 
\begin{figure}[t]
\centering
 \includegraphics[width=1.0\columnwidth]{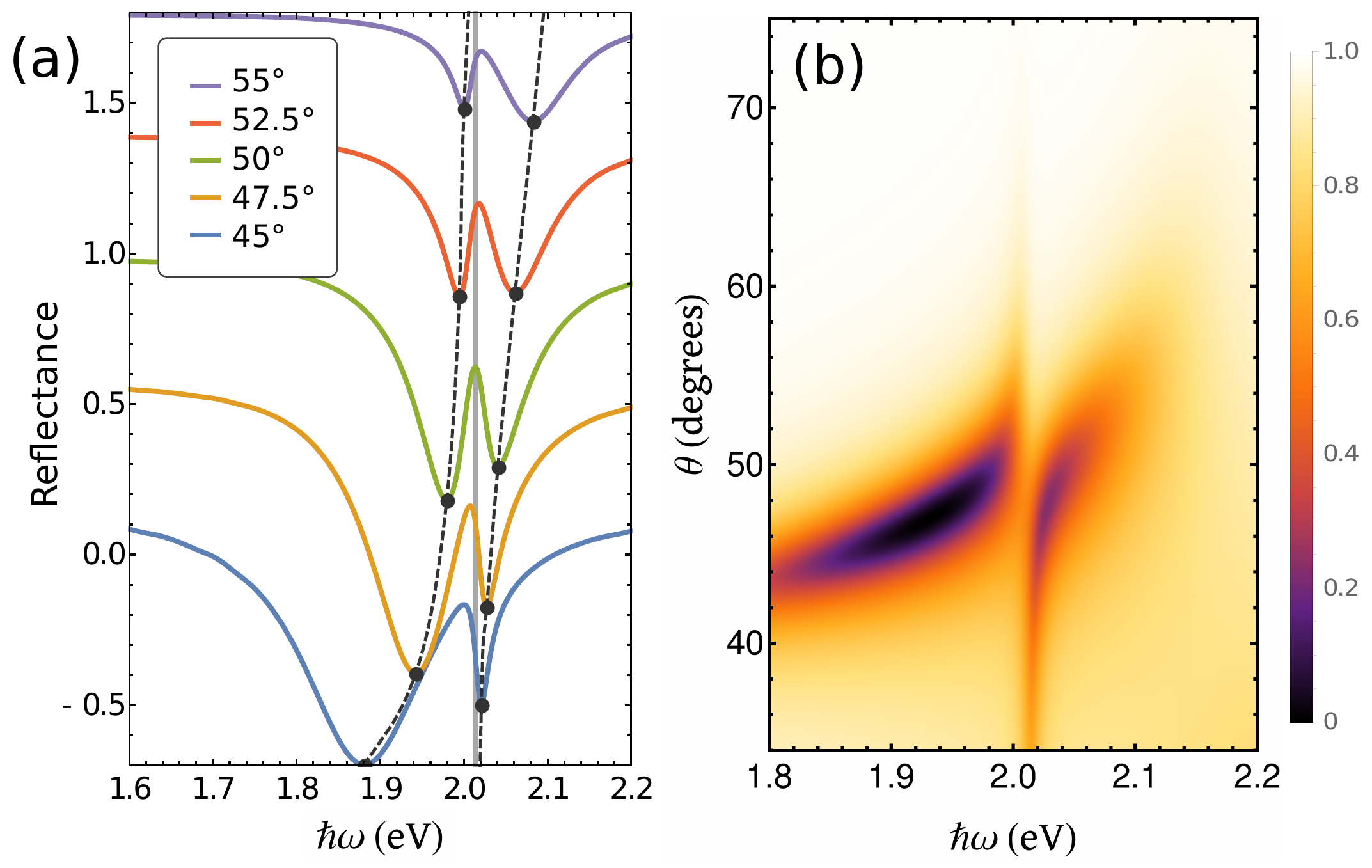}
  \caption{Reflectance spectra in an ATR configuration. 
  (a) ATR spectra collected at different incident angles. 
  The two uppermost (bottommost) curves are shifted vertically by $0.4$ ($-0.4$) each, in the interest of clarity. 
  (b) Two-dimensional plot of the reflectance in an ATR experiment for arbitrary incident angles and excitation energies. 
  The avoided crossing between each polariton branch can bee clearly seen around the A exciton energy at $\hbar\Omega_{\rm ex}=2.014$\,eV. 
  Parameters: $\ep_{\rm d}=5$, $L=90$ nm (dielectric layer thickness), and $\ep_{\rm prism}=16$. We have halved the 
  exciton $\gamma_j$'s used thus far to simulate the effect of reducing the temperature 
  (nevertheless, the general features remain essentially unchanged at room temperature, save for a slight increase in the linewidths).}
  \label{fig:4}
\end{figure}
Figure~\ref{fig:4}(a) depicts the ATR signal that would be obtained in experiments carried out at distinct incident angles. For any of the angles considered, two minima can be readily observed in the ATR spectra, corresponding to the two branches of the plasmon-exciton dispersion relation. Notice that, as we have mentioned above, in spectroscopic measurements at fixed momenta (here given by the angle, $q=\frac{\omega}{c} \sqrt{\ep_{\rm prism}} \sin\theta$), the strong coupling manifests itself in the traditional form of an avoided crossing, together with the transfer of spectral weight between the two polariton branches as the angle is varied---see Fig.~\ref{fig:4}. Moreover, Fig.~\ref{fig:4}(b) illustrates the dependence of the reflectance on both the incident angle and excitation energy. In this regard, each of the curves presented in Fig.~\ref{fig:4}(a) simply correspond to horizontal cuts in Fig.~\ref{fig:4}(b), at the selected angles. In addition, the two minima recognized in the former occur at the energies where such horizontal lines cross each one of the two hybrid plasmon-exciton modes visible in the latter.
We stress the similarities between the spectra portrayed in Fig.~\ref{fig:2} and Fig.~\ref{fig:4}(b) after a reflection with respect to the origin, since these are nothing but two alternative ways to present the polaritonic spectrum.

\begin{figure}[b]
\centering
 \includegraphics[width=1.0\columnwidth]{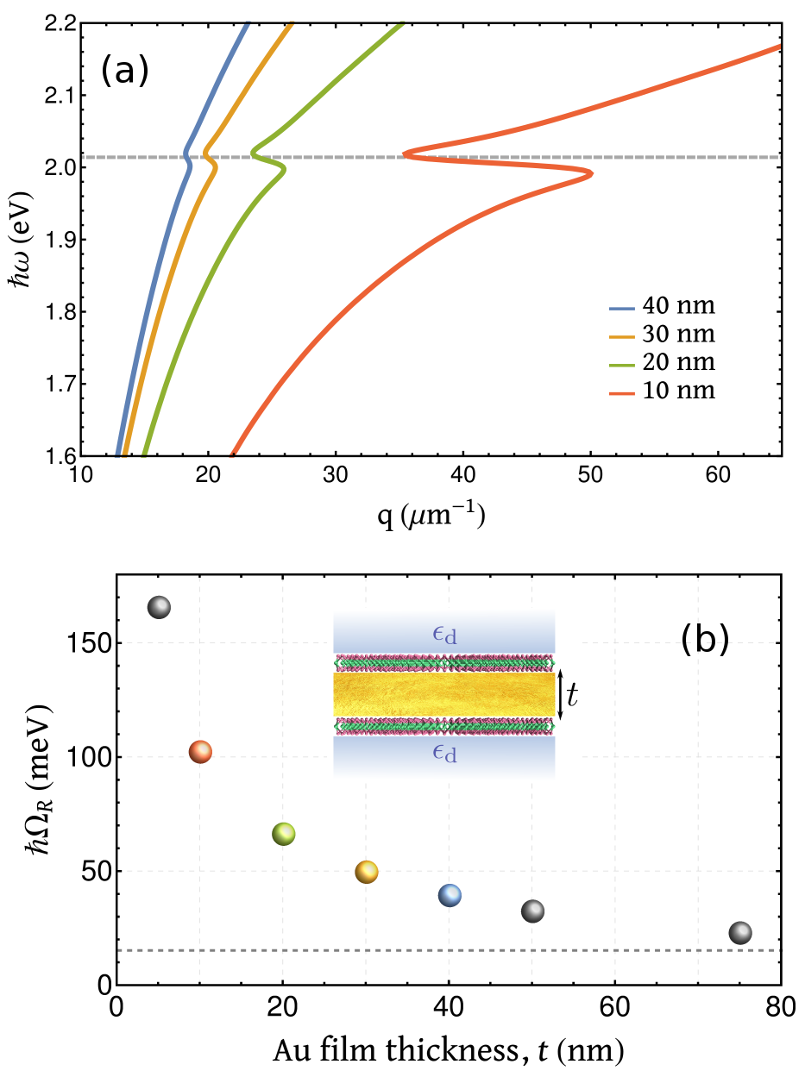}
  \caption{% 
  (a) Dispersion diagram $({\rm Re}\{q\},\hbar\omega)$ of plasmon-exciton polaritons in dielectric/WS$_2$/Au/WS$_2$/dielectric structures with varying film thicknesses. The horizontal dashed line indicates the resonance of the A exciton at $\hbar\Omega_{\rm{ex}}=2.014$ eV. The corresponding Rabi splittings are $\hbar\Omega_R =39,\ 49, \ 66,\ 101$ meV 
  in order of decreasing thickness.
  (b) Rabi splitting energy (in meV) as a function of the Au thin-film thickness 
  in WS$_2$-coated Au films embedded in a homogeneous dielectric medium with $\ep_{\rm d}=2.1$. 
  The colored spheres correspond to the cases plotted in the top panel with matching colors. 
  The horizontal dashed line indicates the $t \rightarrow \infty$ limit (corresponding to a 
  thick Au substrate). 
  In both panels we have used the experimentally obtained optical constants of Au~\cite{ExpAu} 
  and of WS$_2$~\cite{Li:2014}.}
  \label{fig:5}
\end{figure}

\textit{Tuning the plasmon-exciton interaction using thin-films.} 
We further note that an alternative possibility to achieve and tailor the plasmon-exciton 
interaction in TMDC/metal planar structures is to consider a metallic thin-film instead of a semi-infinite metal substrate. 
In such a scheme, the key role played by the dielectric superstrate in tuning the strong coupling is now played by the thickness of the thin-film. In particular, a dielectric/TMDC/metal/TMDC/dielectric structure sustains two modes of opposite symmetries that are thickness dependent. Focusing on the low-frequency mode~\footnote{The other mode with higher frequency tends to the light line as the film thickness is reduced, 
and thus it cannot be used to increase the plasmon-exciton coupling in this structure.}, 
the corresponding spectrum stems from the solution of~\cite{Raza:2013,Goncalves:2016}
\begin{equation}
\tanh(\kappa_{\rm m} t/2) \frac{\epsilon_{\rm m} }{ \kappa_{\rm m}} + \frac{\epsilon_{\rm d}}{\kappa_{\rm d}} = 
\frac{\sigma_{\mathrm{2D}}}{i \omega \epsilon_0} , 
 \label{eq:dispersion_tt}
\end{equation}
where $t$ denotes the thickness of the metallic film. In the above, 
we have assumed a symmetric dielectric environment for the sake of clarity alone. 
Figure~\ref{fig:5}(a) depicts the numerical solution of Eq.~(\ref{eq:dispersion_tt}) 
for a representative structure composed of a WS$_2$-coated gold film embedded in a 
medium with dielectric constant $\ep_{\rm d}=2.1$ (e.g., SiO$_2$). The four curves 
show the dispersion diagrams calculated for thin-film thicknesses varying from 40 nm to 10 nm. 
Clearly, despite the low-index dielectric with just $\ep_{\rm d}=2.1$, large 
Rabi splittings can now be readily achieved by controlling the metal thickness. 
The 
magnitude of the Rabi splitting as a function of the Au film thickness is illustrated in 
Fig.~\ref{fig:5}(b), with the cases plotted in the upper panel highlighted using the same 
colors. Indeed, in those cases, the obtained energy splittings grow from an already significant 
value of 40 meV up to about 100 meV. 
In contrast, using the same dielectric medium to encapsulate a thick gold substrate covered with WS$_2$, as in the initial case, 
one would only obtain a Rabi splitting of about 15 meV [as indicated by the dashed line in Fig.~\ref{fig:5}(b)]. Therefore, these results highlight the benefit of exploiting the thickness-dependent low-frequency SPP mode of metallic thin-films to tailor the plasmon-exciton coupling in planar structures, which are straightforward to fabricate in 
an experimentally convenient fashion.

%
% ==============================================================================================
% ::                                     CONCLUSIONS                                          ::
% ==============================================================================================

\textit{Conclusions.} In conclusion, we have investigated the emergence of plasmon-exciton polaritons in heterostructures consisting in dielectric/TMDC/metal vertical stacks. Throughout the Rapid Communication, we have focused in the case where the TMDC was a WS$_2$ monolayer, thanks to its large exciton oscillator strengths, and where the metal was gold, whose chemical stability has endowed it a popular plasmonic material. Nevertheless, our treatment and subsequent analysis can be straightforwardly generalized for any 2D TMDC/metal pair. Our results plainly demonstrate the emergence of plasmon-exciton interaction within the strong coupling regime upon choosing adequately the dielectric environment enclosing the TMDC/metal structure, or the thickness of the metal film in a double interface configuration. Furthermore, we have shown that the Rabi splitting is enlarged as the relative permittivity of the dielectric layer increases. This behavior can be easily interpreted by noting that for an air/metal interface, the corresponding $\omega_{\rm sp}$ lies well above the exciton resonances 
of TMDCs. Notwithstanding, the placement of a top dielectric layer effectively reduces the nonretarded surface plasmon resonance, thereby successively increasing the spectral overlap between the plasmon and the exciton modes as $\ep_{\rm d}$ is increased. 
Moreover, besides the single-interface case, we have also considered a gold thin-film 
coated with a 2D TMDC and encapsulated in a homogeneous dielectric medium. We have demonstrated 
that this latter configuration constitutes a convenient alternative to the former, since 
the magnitude of the Rabi splitting can be customized upon varying the thin-film thickness, 
which in turn can be controlled experimentally with precision.

Our calculations predict that Rabi splitting energies in excess of $\hbar\Omega_{\rm R}=100$\,meV can be comfortably achieved even at room temperature, which makes our structures viable alternatives to J-aggregates of organic molecules or quantum dots as platforms for studying strong coupling. Remarkably, we have also found that the Rabi splittings stemming from the dielectric/WS$_2$/Au vertical structures considered here are comparable or even surpassing the ones reported in experiments with plasmonic NPs deposited on TMDCs~\cite{Liu:2016,Wen:2017,Abid:2017}, or for TMDCs inside photonic cavities~\cite{liu:2015,flatten:2016}. 
All of this with the added advantage of, in comparison, being extremely easy to fabricate in larger scales. 
Towards an experimentally-oriented study, we have also computed the reflectance measured in a typical ATR experiment, in which plasmon-excitons are excited via prism coupling. Here too, our results---embodied in 
Fig.~\ref{fig:4}---reveal the existence of strong plasmon-exciton interactions within the strong coupling regime. 
It is worth mentioning the resemblance between the spectra exhibited in Fig.~\ref{fig:4}(a) with the spectral features 
akin to strong coupling between plasmons in graphene nanostructures and phonons of the underlying substrate (e.g., hBN or SiO$_2$)~\cite{Atwater:2014,Luxmoore:2014,Goncalves:GNRs}. 
%While the latter occurs in the mid-IR, the hybrid plasmon-exciton modes studied here arise 
%in the near-IR up to the red part of the visible spectrum. 
%However, the underlying physics can be interpreted in a similar way. 

Lastly, it should be mentioned that in our description we have neglected the impact of the dielectric environment surrounding the TMDC on the exciton resonances. 
It is well-known that a landscape with higher permittivity leads to increased screening. The effect is two-fold: it decreases both 
the quasiparticle bandgap and the exciton binding energy, and thus influences the excitonic resonance~\cite{wang2017excitons,raja:2017,Stier:2016,Thygesen:2017}. In previous studies the overall effect 
seems to be small for our purposes (since both effects work against each other)~\cite{raja:2017}, specially in what the A exciton is concerned, and therefore we do not take them into account here. 
However, to what extent that remains valid when the screening is due to an adjacent metal (with much higher permittivity) remains an 
open question, since to the best of our knowledge such a study is still lacking in the literature. 
A possible way to mitigate the impact of the screening by the metal would be to include a thin dielectric layer between the metal and the TMDC. Our calculations show~\cite{SuppMat} that 
the corresponding Rabi splitting in only slightly decreased, and that can be even compensated by increasing $\ep_{\rm d}$ or decreasing $t$.
Nonetheless, we expect that our results and ensuing analysis to be robust against that effect, 
since we anticipate its consequence to be only a shift in the (absolute) position of the excitonic resonances. 
Further, we believe that the simplicity is the main merit of our description, together with its intuitive account of the plasmon-exciton interaction.

We believe that our work contributes to the active pursue of novel plexcitonic architectures based on hybrid 2D/3D structures made from
atomically thin semiconductors and 3D metals. Our findings show that these platforms can be used as testbeds to investigate plasmon-exciton 
interactions in the strong coupling regime, setting the stage for future ultracompact nanophotonic devices operating at room temperature.\\
\hfill

\textit{Acknowledgments.} We thank C. Tserkezis and C. Wolff for 
stimulating discussions. The Center for Nanostructured Graphene is sponsored by the Danish National Research Foundation (Project DNRF103). The Center for Nano Optics is financially supported by the University of Southern Denmark (SDU 2020 funding).
N.~A.~M. is a VILLUM Investigator supported by VILLUM FONDEN (grant No. 16498).

% Bibliography
\bibliography{Plexcitons_2D}

\begin{thebibliography}{34}
\bibitem [{not()}]{noteEps}%
  \BibitemOpen
  \href@noop {} {\ }\bibinfo {note} {The fitting parameters for the Lorentz
  oscillator shown in Fig. 1(b) are: $\ep_\infty=8.76$, $f_j=\{1.9, 0.254,
  0.146, 0.068, 3.07, 1.17, 0.068, 15.5, 12.7\} \mathrm{(eV^2)}$,
  $\hbar\omega_j=\{2.014, 2.185, 2.250, 2.285, 2.402, 2.575, 2.655, 2.845,
  3.047\} \mathrm{(eV)}$, and $\hbar\gamma_j=\{0.029, 0.1, 0.1, 0.1, 0.14,
  0.21, 0.21, 0.265, 0.25\} \mathrm{(eV)}$.}\BibitemShut {Stop}%
\end{thebibliography}

\end{document}